\documentclass[twocolumn]{aa} 
\usepackage{graphicx}
\usepackage{txfonts}
\usepackage{natbib}
\bibpunct{(}{)}{;}{a}{}{,} 

\newcommand{\deltanunu}{\Delta\nu(\nu)}
\newcommand{\dnumoy}{\langle\Delta\nu\rangle}
\newcommand{\HBR}{\mathcal{R}}
\newcommand{\diff}{\mathrm{d}}
\newcommand{\ratednu}{\diff\Delta\nu/\diff n}
\newcommand{\logg}{$\log g$}
\newcommand{\feh}{[Fe/H]}

\newcommand{\kms}{km\,s$^{-1}$}
\newcommand{\maa}{m\AA}
\newcommand{\feone}{Fe\,{\sc i}}
\newcommand{\fetwo}{Fe\,{\sc ii}}
\newcommand{\feonetwo}{Fe\,{\sc i} $+$ {\sc ii}}
\newcommand{\mgoneb}{Mg\,{\sc i}\,b}
\newcommand{\ind}[1]{_{\mathrm{#1}}}
\newcommand{\numaxcor}{\nu\ind{EACF}}
\newcommand{\teff}{$T_{\rm eff}$}
\newcommand{\vmicro}{$v_{\rm micro}$}
\newcommand{\numax}{\nu\ind{max}}

\begin{document}
%
 \title{The solar-like CoRoT\thanks{The CoRoT space mission, launched on December 27th 2006, has been developed and is operated by CNES, with the contribution of Austria, Belgium, Brazil, ESA (RSSD and Science Programme), Germany and  Spain.} target HD~170987: spectroscopic and seismic observations}
 \titlerunning{HD~170987}

   \author{S. Mathur\inst{1,2} \and 
                R.~A. Garc\'\i a\inst{1} \and
                C. Catala \inst{3} \and
                 H. Bruntt\inst{3} \and
                 B. Mosser\inst{3} \and
                T. Appourchaux\inst{4} \and
                J. Ballot\inst{5} \and
                O.~L. Creevey\inst{6,7}\and
                P. Gaulme\inst{4} \and
                S. Hekker\inst{8} \and
		D. Huber\inst{9} \and
		C. Karoff\inst{8} \and
		L. Piau\inst{1}\and
   		C. R\'egulo\inst{6,7} \and
		I.~W. Roxburgh\inst{10,3} \and
		D. Salabert\inst{6,7} \and
		G. A. Verner\inst{10} \and
	         M. Auvergne \inst{3}\and
	         A. Baglin\inst{3} \and
                  W.~J. Chaplin\inst{8} \and         
	         Y. Elsworth\inst{8} \and
	         	E. Michel\inst{3} \and
	         R. Samadi\inst{3} \and
		K. Sato\inst{1} \and
		D. Stello\inst{9}
	}
   \offprints{savita.mathur@gmail.com}
   \institute{Laboratoire AIM, CEA/DSM -- CNRS - Universit\'e Paris Diderot -- IRFU/SAp, 91191 Gif-sur-Yvette Cedex, France
   \and High Altitude Observatory, NCAR, P.O. Box 3000, Boulder, CO 80307, USA
   \and LESIA, UMR8109, Universit\'e Pierre et Marie Curie, Universit\'e Denis Diderot, Obs. de Paris, 92195 Meudon Cedex, France
   \and Institut d'Astrophysique Spatiale, UMR8617, Universit\'e Paris XI, Batiment 121, 91405 Orsay Cedex, France
   \and Laboratoire d'Astrophysique de Toulouse-Tarbes, Universit\'e de Toulouse, CNRS, F-31400, Toulouse, France
   \and Universidad de La Laguna, Dpto de Astrof\'isica, 38206, Tenerife, Spain
   \and Instituto de Astrof\'\i sica de Canarias, 38205, La Laguna, Tenerife, Spain
   \and School of Physics and Astronomy, University of Birmingham, Edgbaston, Birmingham B15 2TT, UK 
   \and Sydney Institute for Astronomy, School of Physics, University of Sydney, NSW 2006, Australia
      \and Astronomy Unit, Queen Mary University of London, Mile End Road, London E1 4NS, UK
   }

   \date{Received 2010; accepted }

 \abstract
     {The CoRoT mission is in its third year of observation and the data from the second long run in the galactic centre direction are being analysed. The solar-like oscillating stars that have been observed up to now have given some interesting results, specially concerning the amplitudes that are lower than predicted. We present here the results from the analysis of the star HD~170987.}
   {The goal of this research work is to characterise the global parameters of HD~170987. We look for global seismic parameters such as the mean large separation, maximum amplitude of the modes, and surface rotation because the signal-to-noise ratio in the observations do not allow us to measure individual modes. We also want to retrieve the stellar parameters of the star and its chemical composition.}
   {We have studied the chemical composition of the star using ground-based observations performed with the NARVAL spectrograph. We have used several methods to calculate the global parameters from the acoustic oscillations based on CoRoT data. The light curve of the star has been interpolated using inpainting algorithms to reduce the effect of data gaps.}
   {We find power excess related to p modes in the range [400 - 1200]\,$\mu$Hz with a mean large separation of 55.2\,$\pm$\,0.8\,$\mu$Hz with a probability above 95\,\% that increases to 55.9 $\pm$\,0.2\,$\mu$Hz in a higher frequency range [500 - 1250] \,$\mu$Hz and a rejection level of 1\,$\%$. A hint of the variation of this quantity with frequency is also found. The rotation period of the star is estimated to be around 4.3 days with an inclination axis of $i$\,=\,$50\degr \;^{+20}_{-13}$. We measure a bolometric amplitude per radial mode in a range [2.4 - 2.9]~ppm around 1000~$\mu$Hz. Finally, using a grid of models, we estimate the stellar mass, M\,=\,1.43~$\pm$\,0.05~$M_\odot$, the radius, R\,=\,1.96~$\pm$\,0.046~$R_\odot$, and the age $\sim$2.4~Gyr. }
  {}
   \keywords{Asteroseismology -- Methods: data analysis --
	     Stars: oscillations -- Stars: individual: HD~170987
	     }

   \maketitle
   
\section{Introduction}

During the present decade the number of confirmed solar-like pulsators -- those with acoustic modes excited by turbulent motions in the near-surface convection \citep[e.g.][and reference therein]{2004SoPh..220..137C} -- has increased enormously thanks, first, to the growing number of ground-based observing campaigns \citep[e.g.][]{2007CoAst.150..106B,2008ApJ...687.1180A}, and second, to the high-precision photometry measurements provided by space instrumentation such as WIRE \citep[Wide-Field Infrared Explorer, e.g.][]{2007A&A...461..619B}, MOST \citep[Microvariability and Oscillations of STars,][]{2003PASP..115.1023W} and CoRoT \citep{2006ESASP.624E..34B}. The latter has been providing data with an unprecedented quality both in terms of photometric precision and in terms of uninterrupted observation lengths. CoRoT has already observed several main-sequence solar-like pulsators \citep{2008Sci...322..558M} while it has enabled to resolve the individual modes of the oscillations spectra of several F stars  \citep{2008A&A...488..705A,2009A&A...506...51B,2009A&A...506...41G,2009A&A...506...33M} and G stars \citep[][Ballot et al. in prep.]{2010arXiv1003.4368D}. At least, it allowed to derive a large spacing for the faintest targets \citep{2009A&A...506...41G,2009A&A...506...33M}. The measurement of these seismic parameters already offer a valuable tool for accurate determinations of radii \citep[e.g.][]{stello09} and ages \citep[e.g.][]{2008arXiv0810.2440C} of stars, which are specially interesting for better understand stellar evolution as well as to characterise stars hosting planets. All of these CoRoT observations are the starting point for a better understanding of the structure \citep{2009A&A...506..175P,2009Ap&SS.tmp..241D} and the surface dynamics \citep{2009A&A...506..167L,2009arXiv0910.4027S,2009arXiv0910.4037S} of this class of stars.

The Kepler mission, successfully launched  in March 7, 2009 \citep{2009IAUS..253..289B}, will also contribute to this field by observing stars on very long runs (4 years). The quality of the first data  on stars showing solar-like oscillations \citep{2010ApJ...713L.176B, 2010ApJ...713L.169C, 2010ApJ...713L.187H, 2010ApJ...713L.182S} promises the asteroseismology a bright future on the study of stellar interiors and dynamical processes \citep[][]{2009arXiv0911.4629C,2009arXiv0912.0817S,2009A&A...506..811M}.

In this paper we present results about a star recently observed by CoRoT, HD~170987 (or  HIP 90851). This target is a well-known double star where components are separated by $0.7\arcsec$ \citep[e.g.][]{2002yCat.1274....0D}. The main star is a F5 dwarf star with a magnitude $m_V$ ranging from 7.4 to 7.7 in the literature, while the second component has a magnitude around 8.5.

\begin{figure*}[!ht]
\begin{center}
\includegraphics[width=15cm, height=6cm, trim=0cm 1.5cm 0cm 9.5cm]{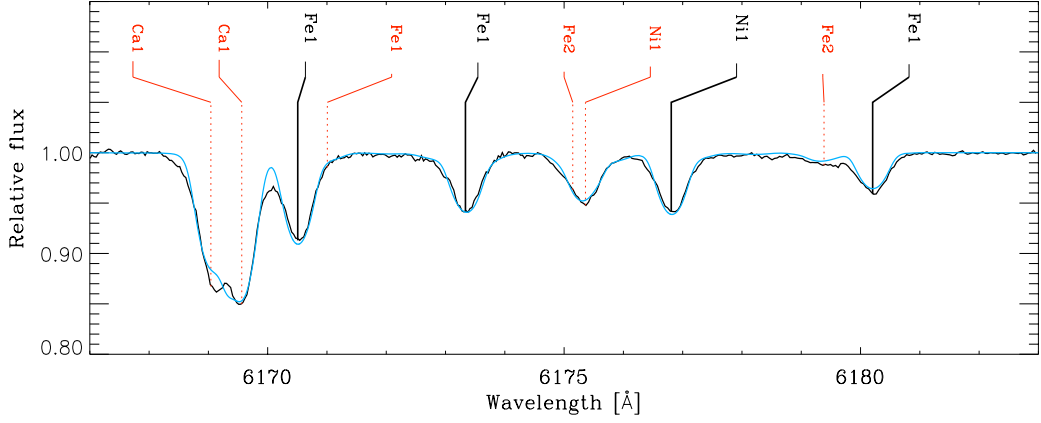}

\caption{Comparison of the observed (black) and computed (blue) spectrum of HD~170987.
The four solid vertical lines mark the lines used in the spectral analysis while the vertical dotted lines show other well-known lines.}
\label{fig:spec}
\end{center}
\end{figure*}

We start by reporting the latest spectroscopic results observed by the NARVAL spectrograph (in Sect.~2), which shows that this star is very similar to Procyon \citep[e.g.][]{allende02}. Then we describe the observations done with CoRoT during 149 days and the interpolation done in the data gaps of the light curve (Sect.~3). In Sect.~4 we infer the surface-rotation period of the star from the detailed analysis of the low-frequency region of the power spectrum and then, we obtain the global properties of the acoustic modes and of the star, respectively in Sect.~5 and Sect.~6.
We finish in Sect.~7 with a discussion of the results and in Sect.~8 with the conclusions of the paper.

\section{Spectroscopic results}\label{secspec}

We have observed HD~170987 using the NARVAL spectrograph 
on the 2-m class Bernard Lyot Telescope at the Pic~du~Midi Observatory.
We acquired one spectrum on each night of 2009 July 7, 8 10, 11 and 13.
The spectra were co-added to obtain a signal-to-noise ratio in the continuum
of ${\rm S/N} \approx 900$ in the range 5800--6400$\,\AA$.
As an example of the quality of the spectrum, a small section is shown in Fig.~\ref{fig:spec}.


We analysed the spectrum using the semi-automatic
software package VWA \citep{bruntt04,2009arXiv0911.2617B}
which uses atmospheric models interpolated in the MARCS grid 
\citep{gustafsson08} and atomic parameters from VALD \citep{kupka99}.
The abundances for 239 lines were calculated 
iteratively by fitting synthetic profiles 
to the observed spectrum using SYNTH \citep{1996A&AS..118..595V}.
For each line the abundances were calculated differentially with respect
to the same line in a solar spectrum from \cite{hinkle00}.
The atmospheric parameters (\teff, \logg, \feh, and \vmicro) were
determined by requiring that Fe lines gave the same abundance 
independent of equivalent width (EW), excitation potential (EP) or
ionization stage. Only weak lines were used (${\rm EW}<90$\,\maa) for
this part of the analysis as they are sensitive to adjust the parameters, while stronger lines (${\rm EW}<140$\,\maa) were
used for the calculation of the final mean abundances. 
The uncertainties on the atmospheric parameters and the abundances 
were determined by perturbing the best-fitting atmospheric 
parameters, as described by \cite{bruntt08}.

\begin{figure}[!htb]
\begin{center}
\includegraphics[width=10cm, trim=3cm 15cm 0cm 2cm]{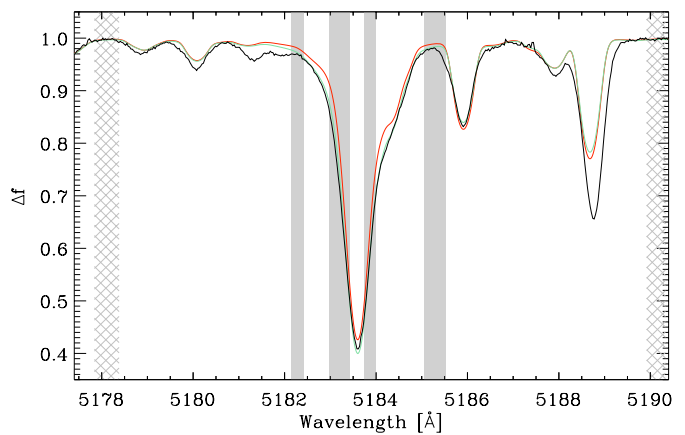}
\caption{Two synthetic profiles with different surface gravity
are compared to the observed (black) spectrum ($\log g = 4.35$ (green line) and $3.95$ dex (red line)).
The hatched regions are used for renormalizing the spectrum.
The solid rectangles mark the regions where
the $\chi^2$ is computed to determine the optimal value of $\log g$.}
\label{fig:mg}
\end{center}
\end{figure}

\begin{figure*}[!htb]
\begin{center}
\includegraphics[width=18cm, trim= 0 0 0 12cm]{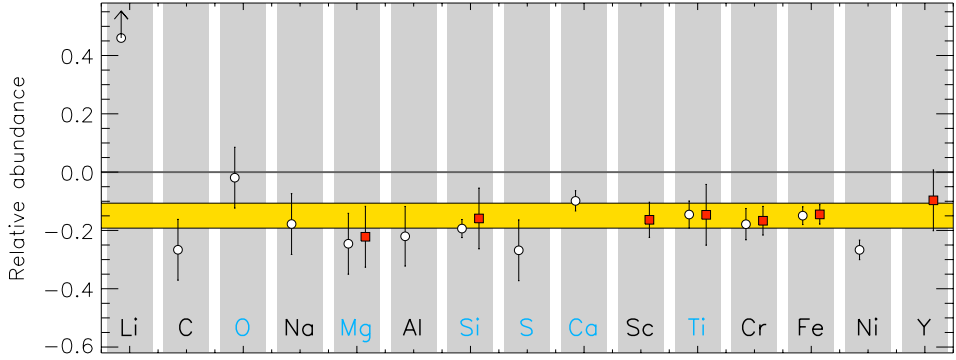}
\caption{Abundance pattern of HD~170987 for 16 elements. Circles and box symbols
are used for the mean abundance from neutral and singly ionised lines,
respectively. Notice the high abundance of lithium. The yellow horizontal bar marks the mean metalicity with 1-$\sigma$ uncertainty range.}
\label{fig:abund}
\end{center}
\end{figure*}

For F-type stars like HD~170789, non-LTE (local thermodynamical equilibrium) effects cannot be ignored.
Our model of atmospheres assumes LTE and we therefore 
systematically underestimate the abundance of \feone\ while \fetwo\ 
is nearly unaffected. We followed the approach of \cite{bruntt08} and 
corrected the \feone\ abundance using the results of \citet{1996A&A...312..966R}. 
The effect is [Fe\,{\sc i}/H]$_{\rm NLTE} = $ [Fe\,{\sc i}/H]$_{\rm LTE}$ + 0.046$\,$dex.
This affects our estimate of the surface gravity \logg , as was discussed
by \cite{fuhrmann97} in the case of Procyon, which has a very similar spectral type.
Following their approach we used the \mgoneb\ lines at 5173 and 5184\,\AA\
to determine \logg. We first adjusted the Van der Waals broadening 
constants from VALD in order to fit the solar spectrum for the canonical
value of $\log g = 4.437$ dex. We then determined the Mg abundance in HD~170987 from two
weak Mg lines and fitted the wings of the two strong \mgoneb\ lines. 
An example of the fit is shown in Fig.~\ref{fig:mg}, where the observed spectrum 
is compared to two synthetic spectra with $\log g = 4.35$ and 3.95~dex. 
The hatched regions near the border of the plot are used to normalize the spectrum and the grey-shaded
rectangles show the sections of the spectrum used to compute the $\chi^2$ value needed to
determine the best value of \logg. The mean value of the two \mgoneb\
lines gives $\log g = 4.35\,\pm\,0.22$~dex. This value is slightly 
higher but in agreement with the more precise value determined from 
the analysis of \feonetwo\ lines: $\log g = 4.20\,\pm\,0.05$~dex. 
Although the latter analysis is affected by our adopted correction
for the NLTE effect, we use it as our final estimate of the surface gravity.

From our analysis of the spectrum of HD~170987 we determine the following parameters:
$T_{\rm eff} =  6540~\pm\,36 \pm 70$\,K,
$\log g = 4.20 \pm 0.05 \pm 0.10$ dex,
 and $v_{\rm micro} = 1.70\,\pm\,0.10\,\pm\,0.05$\,\kms.
We give two uncertainties for these values, the first being the intrinsic error
and the second being the estimated systematic error. The systematic errors
were estimated from a large sample of F5--K1 type stars (Bruntt et al., in prep.). 
These two uncertainties must be added quadratically (e.g.\ $T_{\rm eff} = 6540\,\pm 80$\,K).
For the mean metallicity we use the \fetwo\ lines to get [M/H]=$-0.15\,\pm\,0.06$ dex.
In Table~\ref{tab:abund} we list the mean abundances of 16 elements compared to the Sun \citep{2007SSRv..130..105G},
which are also shown in Fig.~\ref{fig:abund}. Finally, for the projected
rotational velocity we find v~$\sin i = 19.0\,\pm\,1.5$\,\kms, which is significantly greater than for Procyon.

We have also generated a mean profile from the spectrum using the Least-Squares Deconvolution Technique developed by \citet{1997MNRAS.291..658D}. This shows evidence that the spectral lines are slightly asymmetric, indicating that the spectrum is contaminated by a fainter star. The effect on the determined spectroscopic parameters are only weakly affected by this.

\begin{table}
 \centering
 \caption{Chemical composition of HD~170987 measured relative to the Sun ($\Delta A$). $n$ is the number of spectra lines used.
 \label{tab:abund}}
\begin{tabular}{l|lr|lllr}
\hline
\hline
El.            & $\Delta A $ (dex)            & $n$  & El.             & $\Delta A$   (dex)         & $n$ \\
\hline

  {Li \sc   i} &     $  1.80       $   &   1  &    {Ca \sc   i} &     $ -0.10\,\pm\,0.03$   &  11  \\ 
  {C  \sc   i} &     $ -0.27       $   &   2  &    {Sc \sc  ii} &     $ -0.16\,\pm\,0.06$   &   3  \\ 
  {O  \sc   i} &     $ -0.02       $   &   1  &    {Ti \sc   i} &     $ -0.15\,\pm\,0.05$   &   4  \\ 
  {Na \sc   i} &     $ -0.18       $   &   2  &    {Ti \sc  ii} &     $ -0.15\,\pm\,0.08$   &   2  \\ 
  {Mg \sc   i} &     $ -0.25       $   &   1  &    {Cr \sc   i} &     $ -0.18\,\pm\,0.05$   &   7  \\ 
  {Mg \sc  ii} &     $ -0.22       $   &   1  &    {Cr \sc  ii} &     $ -0.17\,\pm\,0.05$   &   5  \\ 
  {Al \sc   i} &     $ -0.22\,\pm\,0.10$   &   4  &    {Fe \sc   i} &     $ -0.15\,\pm\,0.03$   & 143  \\ 
  {Si \sc   i} &     $ -0.19\,\pm\,0.03$   &  11  &    {Fe \sc  ii} &     $ -0.15\,\pm\,0.03$   &  20  \\ 
  {Si \sc  ii} &     $ -0.16\,\pm\,0.03$   &   2  &    {Ni \sc   i} &     $ -0.27\,\pm\,0.03$   &  16  \\ 
  {S  \sc   i} &     $ -0.27       $   &   2  &    {Y  \sc  ii} &     $ -0.10       $   &   1  \\ 
  \hline
  \hline
\end{tabular}
\end{table}

In the catalog of the Geneva-Copenhagen Survey of Solar neighbourhood III \citep{2007A&A...475..519H}, based on Str\"omgren photometry, a value of $m_V$=7.46 was given. The new reduction of Hipparcos data \citep{2007ASSL..350.....V,2007A&A...474..653V} gives a parallax of  $\pi$= 11.21\,$\pm$\,1.02 mas, corresponding to a distance of 89 $\pm$\,9 pc, leading to an absolute magnitude of  $M_V$= 2.71. 

To get a first estimate of the seismic parameters of this star, we can use the value for the effective temperature found. 
The Hipparcos parallax combined with a bolometric correction of -0.01 $\pm$\,0.01 derived using $T_{\rm{eff}}$ in the calibration of \citet{flower96} gives $L/L_{\sun}$ = 6.41\,$\pm$ 1.2 (neglecting interstellar reddening) and consequently $R/R_{\sun}$ = 2.1~$\pm$\,0.2. A rough comparison with solar-metallicity evolutionary tracks gives a mass of $M/M_{\sun}$ = 1.4~$\pm$\,0.2, which finally yields an estimate of the position of maximum amplitude, $\nu\ind{max}$ = 910\,$\pm$\,50\,$\mu$Hz and the mean large separation, $\Delta\nu$\,$\sim$\,52.7~$\pm$ 4\,$\mu$Hz \citep{kjeldsen95}. Compared to the other published CoRoT F stars \citep[e.g.][]{2009A&A...506...51B,2009A&A...506...41G}, HD~170987 is therefore notably more similar to Procyon \citep[e.g.][and references therein]{2008ApJ...687.1180A}.


\section{CoRoT photometric observations}
\begin{figure}[!htb]
\begin{center}
\includegraphics[width=0.47\textwidth, trim=2cm 2cm 0 0cm]{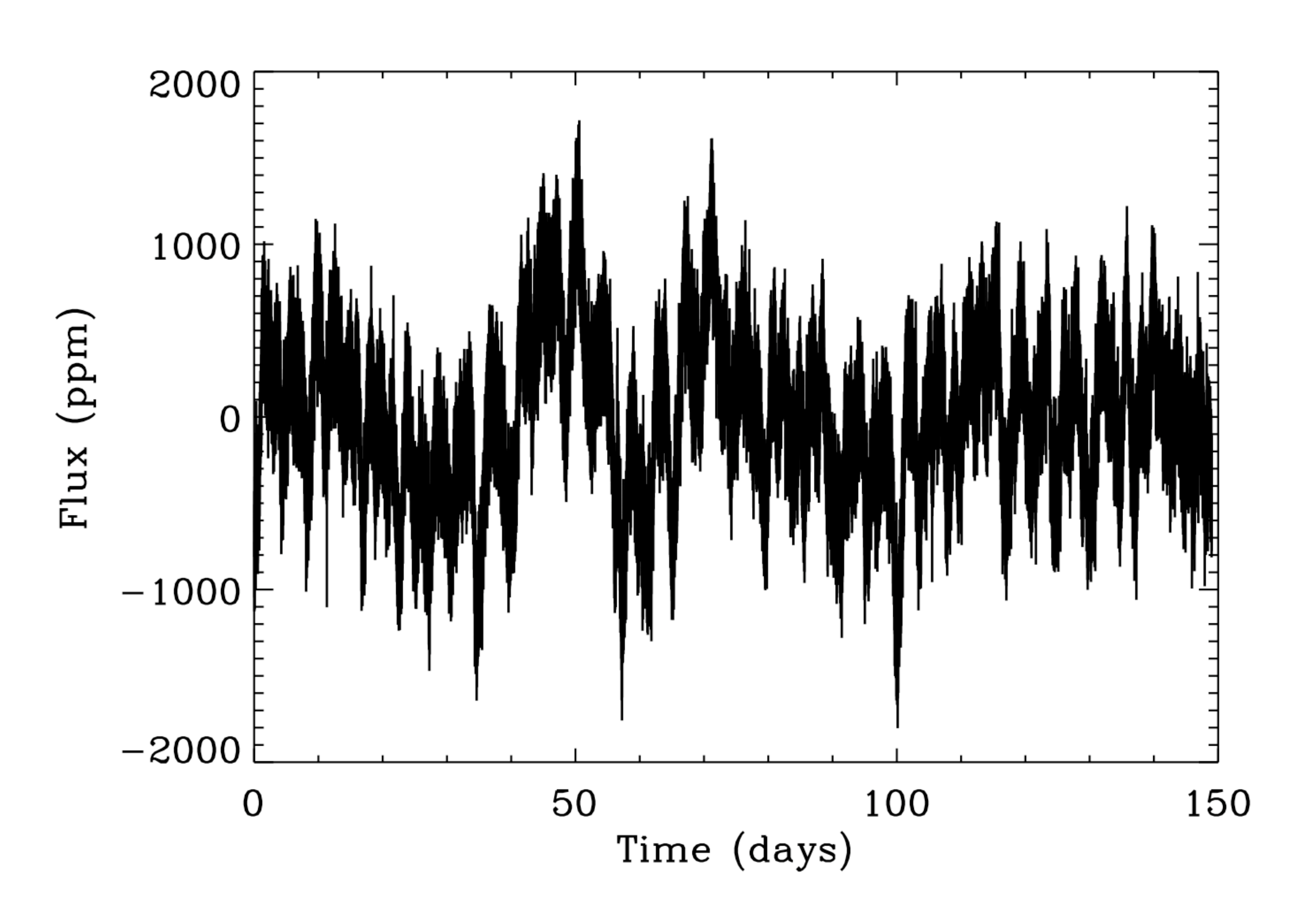}
\caption{N2-Helreg time series of the CoRoT target HD 170987 corrected from the ageing of the CCD. For plotting purposes only every fifth point is shown. }
\label{fig:lcurve}
\end{center}
\end{figure}

HD~170987 has been observed by CoRoT in the seismic field during 149 days starting on April 11, 2008 till September, 7, 2008, during the second long run in the galactic centre direction (LRc02). In Fig.~\ref{fig:lcurve} we show the N2 light curve (in ppm) corrected for the ageing of the CCD by dividing by a third-order polynomial fit. We use time series that are regularly spaced in the heliocentric frame \citep[i.e. the so-called Helreg level 2 data, ][]{2007astro.ph..3354S} with a 32~s sampling rate. The overall duty cycle of the time series is 89.7$\%$.

\begin{figure}[!htb]
\begin{center}
\includegraphics[width=0.47\textwidth, trim=2.5cm 1.5cm 4cm 1.5cm]{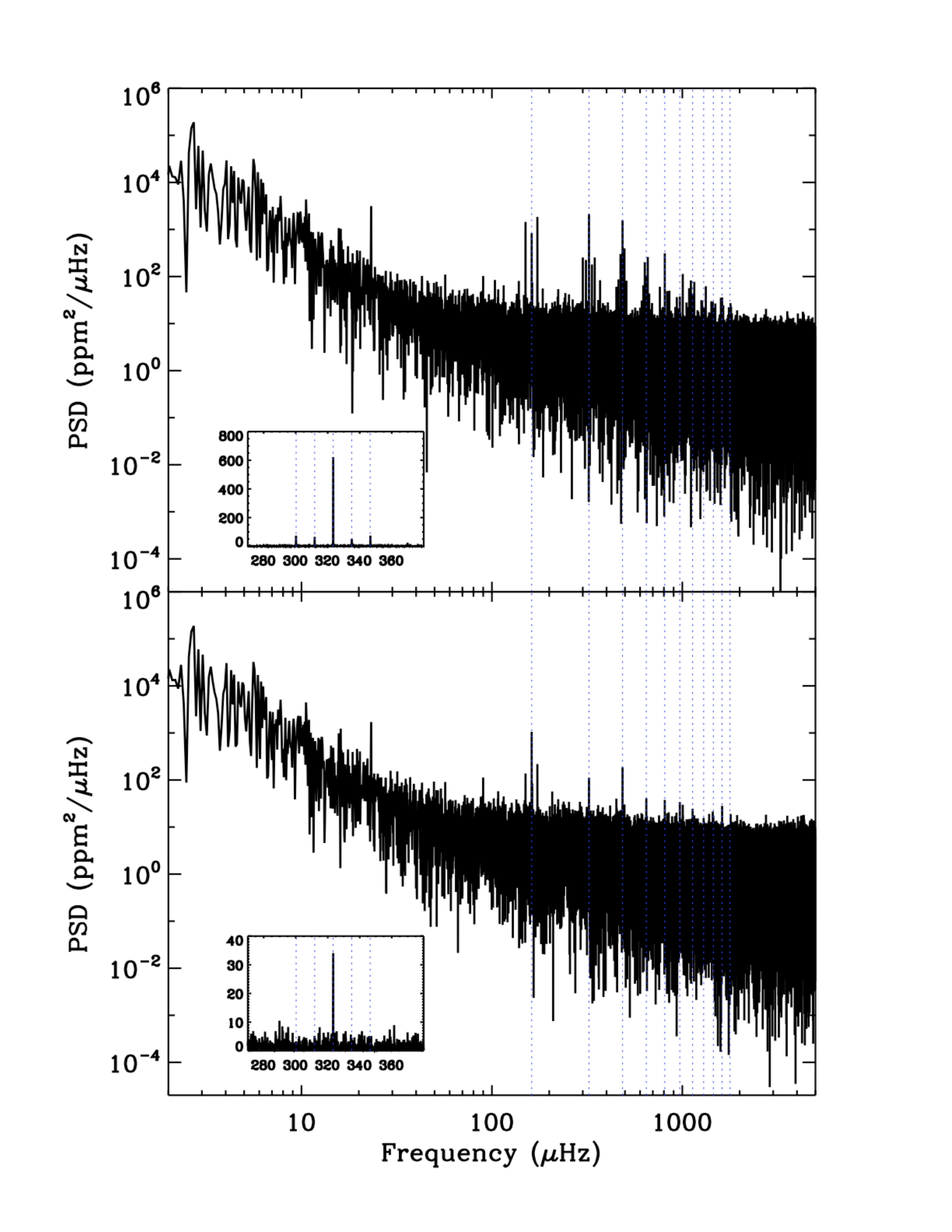} 

\caption{PSD (in units of ppm$^2$/$\mu$Hz) of the raw N2-Helreg time series (top) and after interpolating the data gaps by a Multi-Scale Discrete Cosine Transform as explained in the text (bottom). The vertical dotted lines represent the frequency of the CoRoT orbit at 161.7 $\mu$Hz and the first 10 harmonics. The inserts are a zoom around the second orbital harmonic, normalized by its standard deviation, showing daily modulation with peaks at $\pm$\,11.57 $\mu$Hz and twice this frequency.  }
\label{psdfull}
\end{center}
\end{figure}

To compute the power spectrum density (PSD) we used a standard fast Fourier transform algorithm and we normalized it as the so-called one-sided power spectral density \citep{1992nrfa.book.....P}. The resulting PSD is shown in Fig.~\ref{psdfull}.

The existing gaps are mostly due to the crossing of the South Atlantic Anomaly (SAA) \citep[see for more details][]{2009A&A...506..411A}. The influence of the perturbations of SAA crossings is not completely attenuated by
the standard correction procedure and
thus yields alias peaks at multiples of 
the CoRoT orbital frequency of 161.7 $\mu$Hz (vertical dotted lines in Fig.~\ref{psdfull} (top)). For each of them a daily modulation can be observed at $\pm$\,11.57 $\mu$Hz and twice this frequency (see the inset of Fig.~\ref{psdfull} (top)). In particular, the region where we expect to find the acoustic-mode power excess (centered around 900 $\mu$Hz) is completely dominated by this peak structure. Thus, we need to interpolate the data in the best way to reduce these effects as much as possible as it was already done in HD~175726 \citep{2009A&A...506...33M}. For that star the algorithm proposed by \citep{Holds89} was used giving excellent results. In our case, we adopted a different strategy based on the ``inpainting'' technique using a Multi-Scale Discrete CosineTransform (see Appendix A for a more detailed explanation on the algorithm). This algorithm has been tested and the results have been verified \citep[see for further details][]{2010arXiv1003.5178S}. Moreover, different groups have also used other interpolation techniques (linear, based on wavelets, etc) leading to results for the global parameters of the modes (such as the mean large separation) that agree within the error bars.

\section{Surface rotation}\label{sectrot}

An independent measurement of the surface rotation can be obtained through the signature left by the active regions crossing the visible stellar disk. It provides a first estimation of the stellar rotation. Moreover, combined with the measured v~$\sin i = 19.0\,\pm\,1.5$~\kms (see Sect.~2) and assuming a given radius of the star, we can also estimate the inclination axis of the star, which can be very helpful to better constrain the fitting techniques to characterise the acoustic modes 
\citep{2003ApJ...589.1009G,2006MNRAS.369.1281B,2008A&A...486..867B}.

\begin{figure}[!htb]		
\includegraphics[width=0.47\textwidth, trim=1cm 2cm 1cm 3cm]{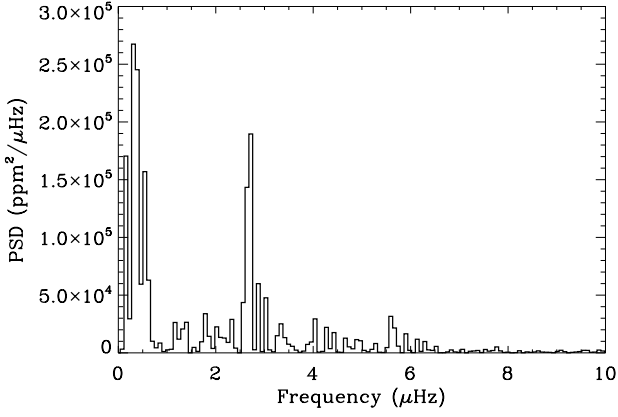}
\caption{Zoom on the low-frequency part of the PSD. The spectrum has been computed with a classical FFT using the full time series.}
\label{figt}
\end{figure}

The light curve of HD~170987 shows a periodic modulation of about 4 days (see Fig.~\ref{fig:lcurve}), which could be due to the combination of the stellar magnetic activity (star spots) and the rotation of the star. It is interesting to analyse this periodicity in a more detailed way by studying the low-frequency range of its power spectrum. This is shown in Fig.~\ref{figt} where a clear peak appears at 2.70 $\mu$Hz. This frequency corresponds to a period of 4.3 days, which can be associated to the rotation period of the surface of the star. 

At lower frequency --around 0.32 $\mu$Hz-- another structure appears which could be related to low-frequency noise. Indeed, considering that the radius is estimated to $\sim$~2~$R_\odot$ and that the observed v~$\sin i$ is of $\sim$~19\,\kms, the second peak is very unlikely to be produced by the rotation of the stellar surface.

Spot modelling of the light curve was performed according to the method exposed in \citet{2009A&A...506..245M}. Since all the stars analyzed in that paper, with a spectral type close to HD 170987, show evidence of differential rotation, we used the fitting method D (Table 2 of \citeauthor{2009A&A...506..245M} \citeyear{2009A&A...506..245M}) with a fixed value of the differential rotation profile. This method has proven to be efficient for the determination of the rotation period and of the spot lifetime. Measurements of the star inclination and differential rotation are much less precise.

\begin{figure}[!htb]		
\includegraphics[width=0.50\textwidth, trim=1.5cm 13cm 4cm 6cm]{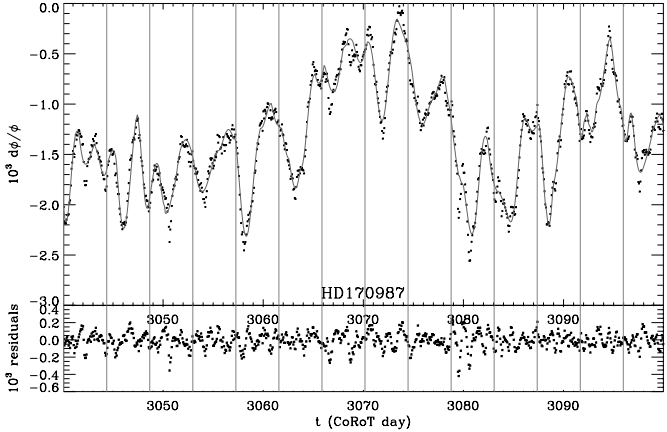}
\caption{Spot modelling of HD 170987. The dots in the upper panel represent the data binned every CoRoT orbit (6184 s) and the solid curve is the best fit model. Dots in the bottom panel represent residuals. Vertical grey lines indicate the mean rotation period.}
\label{taches}
\end{figure}

The best fit model is presented in Fig.~\ref{taches} for a 60-day long sample.  Like similar type of stars, the modelling is not able to reproduce the sharpest features of the light curve. Fitting them would require the introduction of too many spots, which lowers the global precision of the parameters of the best fit. This fit is obtained for an inclination of $40^{+20}_{-10}{}^\circ$. The equatorial rotation period is $4.30\pm 0.05$\, days and the differential rotation rate $5^{+10}_{-5}$\,\%, for a  mean rotational period close to 4.40\,d.  The spot lifetime, 2.5$^{+1.0}_{-0.5}$\,days is significantly shorter than one rotation period of the star.

\begin{figure}[!htb]
\begin{center}
\includegraphics[width=9cm, trim=3cm 90 30 90]{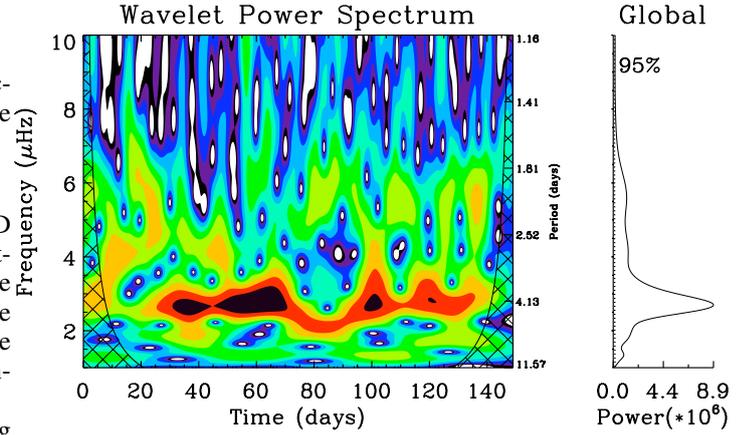}

\caption{Left panel: Wavelet power spectrum for HD~170987 as a function of the frequency of the Morlet wavelet (y-axis) and time (x-axis). The black grid represents the cone of influence. Right panel: Global power spectrum as a function of the frequency of the wavelet. The dotted line is the confidence level corresponding to a 95\% probability.}
\label{wavelet}
\end{center}
\end{figure}

We have used the wavelet technique to confirm the rotation period of the surface of the star \citep{1998BAMS...79...61T}. We have used the Morlet wavelet (a moving Gaussian envelope with a varying width) to produce the wavelet power spectrum shown in Fig.~\ref{wavelet}. This spectrum is a representation of the correlation between the wavelet with a given frequency (y-axis) along time (x-axis). As showed in \citet{2010A&A...511A..46M}, the strength of this method is to see the evolution with time of the power, as well as to resolve the uncertainty between the fundamental period and the first harmonic that could happen, as in the solar case \citep{2008arXiv0810.1803M}. The left panel of the plot shows the presence of power in the wavelet spectrum for periods between 4 and 5 days. This power excess is observed all along the 149 days of observation. When collapsed over the observation time (right panel) we see the accumulation of power at 4.3 days well above the 95$\%$ confidence level limit. 

The reliability of the result depends on two factors: the cone of influence, which is related to the fact that the periodicity has to be at least one quarter of the total length of the time-series and a confidence level corresponding to a 95\% of probability that the peak observed is not due to noise. Thus, using this methodology we determine that the rotation period of HD~170987 is $\sim$4.3 days (or 2.7~$\mu$Hz), which is compatible with the analysis of the low-frequency PSD described above.

It is very clear that the power present around 2.7 $\mu$Hz is concentrated between the 30th and the 70th day (black shaded region) as well as during another small period around the 100th day. This agrees with the results obtained with the PSD.

Given the value of the rotation period and the v~$\sin i$, we can deduce the inclination angle $i$ = $50\degr \;^{+20}_{-13}$.

\section{Global parameters of the acoustic modes}

Using different methods, we have estimated the global parameters of the acoustic modes of HD~170987, such as the location of the p-mode bump, the mean large spacing, and the maximum bolometric amplitude per radial mode. 


\subsection{P-mode excess power in the PSD}\label{sectback}
 
\begin{figure}[htbp]
\begin{center}
\includegraphics[width=9cm, trim=3cm 2cm 1cm 3cm]{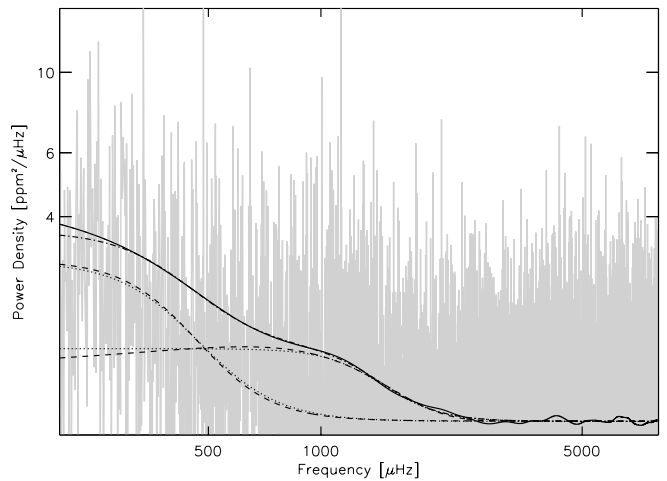}

\caption {Smoothed PSD (continuous solid line) overlaid by the background fitted model assuming two Harvey-model components (dotted lines) and a single Harvey model plus a p-mode excess (dashed lines). The individual components of the fitted background are also shown. The PSD have been smoothed with a Gaussian running mean with a width of 200 $\mu$Hz.}
\label{back}
\end{center}
\end{figure}

To evaluate the possible presence of p-mode oscillations in the spectrum we start by analyzing a smoothed version of the PSD. We smoothed the PSD in the range 200 to 8000 $\mu$Hz with a Gaussian running mean with a width of 200 $\mu$Hz. The reason not to include the low-frequency part of the spectrum is that this region of the PSD mainly bears the signatures of the slow trends in the data and of stellar activity (rotation), which is already explained in Sect.~\ref{sectrot}. The smoothed PSD is shown in Fig.~\ref{back}. Two clear bumps are seen in the spectrum, one around 300 and one around 1000 $\mu$Hz. The bump around 1000 $\mu$Hz could originate from the p-mode oscillations (as it corresponds to the expected $\nu\ind{\rm max}$) but it could also be a signal related to granulation. Up to now, the background of all the precedent CoRoT solar-like targets (HD~49933, HD~175726, HD~181420, HD~181906 and HD~49385) has been correctly characterized by a single component in the Harvey model \citep{1985ESASP.235..199H} that has been associated to granulation. However, when comparing the spectrum of HD~170987 to the one of the Sun obtained using the VIRGO instrument \citep{1995SoPh..162..101F} and to solar-type stars observed by the Kepler satellite (Chaplin et al. 2010) we see that in these cases a second Harvey power law has been used to take into account some extra power located around 1000 $\mu$Hz (time-scale close to 80~s). Based on the smoothed PSD we can therefore not conclude that the spectrum shows p-mode oscillations while we see some excess power centered at $\sim$ 1000$\,\mu$Hz.

For the precise analysis of the acoustic background, different strategies have been followed to study the bump at 1000 $\mu$Hz. 
We start by assuming that the bump around 300\,$\mu$Hz originates from granulation and the one at $\sim$~1000$\,\mu$Hz could originate from a shorter convective scale or by faculae. 

The faculaes are the bright points seen on visual  solar images often close to the dark sunspots. They are due to changes in the opacity caused by a strong magnetic field which means that we see the inside rather than the surface of the granulation and as the temperature is higher inside the granulation cells than at their surfaces the faculaes appear brighter than there surroundings \citep{2004ApJ...607L..59K}. The carateristic timescale of the faculae is shorter than the time-scale of the granulation cells because the faculaes only sample the edges of the granulation cells.

Following Harvey (1985) we model the smoothed PSD in the range [200 - 8000]\, $\mu$Hz with a background model containing two components plus the photon-noise contribution ($c$) that dominate the PSD at high frequency, i.e.

\begin{equation}
PSD(\nu)=c+\frac{4\sigma\ind{gran}^2\tau\ind{gran}}{1+(2\pi\tau\ind{gran}\nu)^{a\ind{gran}}}+\frac{4\sigma\ind{facu}^2\tau\ind{facu}}{1+(2\pi\tau\ind{facu}\nu)^{a\ind{facu}}}.
\end{equation}
Here, $\sigma\ind{gran}$ and $\sigma\ind{facu}$ are the amplitudes of the background signal of granulation and the second convective component or faculae respectively, $\tau$ is the time-scale, and $a_{\rm gran}$ and $a_{\rm facu}$ are two constants. Because we are modelling a smoothed version of the PSD we can assume that the error between the model and the observations are normally distributed. We can therefore fit the model to the observations by means of least squares. Using the robust non-linear least squares curve fitting IDL package MPFIT (http://www.physics.wisc.edu/$\sim$craigm/idl/idl.html) we obtain a granulation time-scale of 383\,$\pm$\,28~s and a second component with a time-scale of 113\,$\pm$\,51~s.

We then assumed that the second bump, around 1000~$\mu$Hz, is caused by p modes. Thus, for the background fitting, we use the first two terms of equation (2) -- the photon noise and one Harvey model component-- and fitted a Gaussian function to the second bump. This led to a very similar fit compared to the previous one, with the same parameters for the granulation time-scale (see Fig.~\ref{back}).  Consequently, both assumptions could be correct and we cannot disentangle which one is the best one at this stage of the analysis.

\subsection{Estimating the mean large separation}







In order to determine whether HD 170987 presents solar-like oscillations, we have applied several methods. One of them consists of searching for the signature of the mean large separation of a solar-like oscillating signal in the autocorrelation of the time series. As proposed by \cite{2006MNRAS.369.1491R}, in a refined application of the Wiener-Khinchine theorem, this autocorrelation is calculated as the Fourier spectrum of the windowed Fourier spectrum. \citet{2009A&A...506..435R} subsequently used a narrower window to look for the variation of the large separation with frequency. 

\cite{2009A&A...508..877M} have shown how to optimize the method and to determine its reliability. They have scaled the autocorrelation function according to the noise contribution, in order to statistically test its significance. Then, when the envelope autocorrelation function (EACF) gives a signal above a defined threshold level, the null hypothesis can be rejected, and a reliable large separation can be derived. For a blind analysis of the mean value, $\dnumoy$, of the large separation and a rejection of the null hypothesis at the level 1\,\%, the threshold level is fixed to 8. This mean value is determined in a frequency range centered on the frequency $\numaxcor$ where the EACF reaches its maximum amplitude, with a filter of width equals to the full width at half-maximum of the mode envelope. Note that $\numaxcor$ is close to $\numax$, where the mode envelope reaches its maximum amplitude, but they are not equal. The difference between $\numaxcor$ and $\numax$ is due to the varying mode lifetime.

The method has shown to be efficient in low signal-to-noise cases, such as the CoRoT target HD 175726 \citep{2009A&A...506...33M} or in the K1V target with a height-to-background ratio $\HBR$ as low as 2.5\,\% \citep{2009A&A...506....7G}. The ratio $\HBR$ measures the maximum smoothed height of the modes compared to the background. HD 170987 is also a challenging target, with $\HBR \simeq 4\,$\%. For HD 170987, with an optimized filter of width 580\,$\mu$Hz, the EACF peaks at the frequency 930\,$\mu$Hz, at a value of 19.2, and gives a mean large separation of $55.9\,\pm\,0.2$\,$\mu$Hz. 


Other methods have been used giving a mean large separation between 54 and 56~$\mu$Hz  \citep{2010MNRAS.402.2049H,2010A&A...511A..46M}. 

\begin{figure}
\centering
\includegraphics[width=9cm, trim=4cm 0 0cm 2cm]{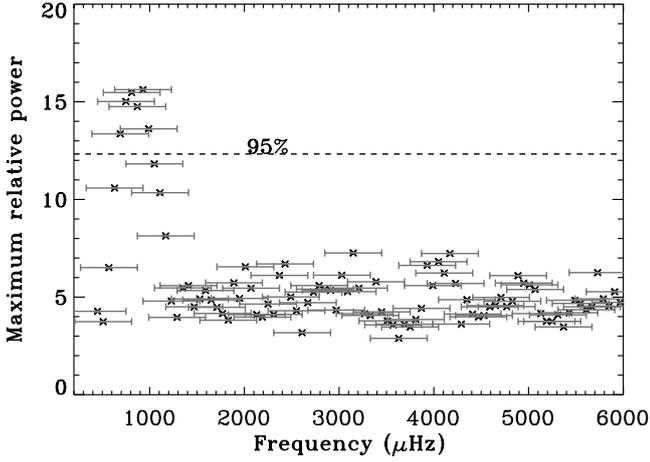}
\caption{Maximum relative power of the highest peak in the PS2 around half of the large separation, as a function of the central frequency of the sliding box taken in the PSD. The dashed line represents the threshold corresponding to a 95\% confidence level. The horizontal error bars represent the boxes in which the PSD has been sliced (600~$\mu$Hz in this case).}
\label{freq_range}
\end{figure}

Indeed, using the method described in \citet{2010A&A...511A..46M}, we have calculated the power spectrum (PS) of the power spectrum (PS2) between 100 and 10000~$\mu$Hz. We select the highest peak and we assume that it corresponds to half of the mean large separation. Then, we cut the original PSD in boxes of 600~$\mu$Hz, shifted every 60~$\mu$Hz and we compute the PS2 of each box looking for the highest peak close to the value found originally. In Fig.~\ref{freq_range} we present the power of the highest peak found in each box (normalised by the $\sigma$ of the PS2).  The horizontal error bars mark the frequency range covered by each box. 
We have obtained that the frequency-range of the p modes is [400 - 1200]~$\mu$Hz with $\dnumoy$~=~55.2~$\pm$\,0.8~$\mu$Hz with 95\% confidence level.



Finally, we took into account that $\Delta \nu$ depends on frequency and we computed the PS2 of a power spectrum with a stretched frequency axis \citep{2010MNRAS.402.2049H}. The stretching is performed in such a way as to produce an equidistant pattern of peaks on the stretched, as opposed to the original, frequency axis. It is done by applying a quadratic correction to the axis values. This allows us to measure the linear variation of the $\Delta\nu$ with frequency (which are in themselves first differences of frequencies). The PS2 of the stretched power spectrum will therefore show a stronger (more prominent) signature of $\Delta \nu$ than the PS2 of the original spectrum. In the PS2 we determine the position of the $\Delta \nu$/2 and $\Delta \nu$/4 features, and compute the Bayesian posterior probability of the features being due to noise. Using these probabilities we then compute the posterior weighted centroids of the features. The interval with a probability of the feature not being due to noise higher than 68.27\%, i.e., 1$\sigma$ in a Gaussian distribution, is used as the uncertainty interval. For this star we could only find a $\Delta \nu$/2 feature in the PS2, which was used to determine $\langle \Delta \nu \rangle$. 

\begin{figure*}[!htb]
\centering
\includegraphics[width=9.cm]{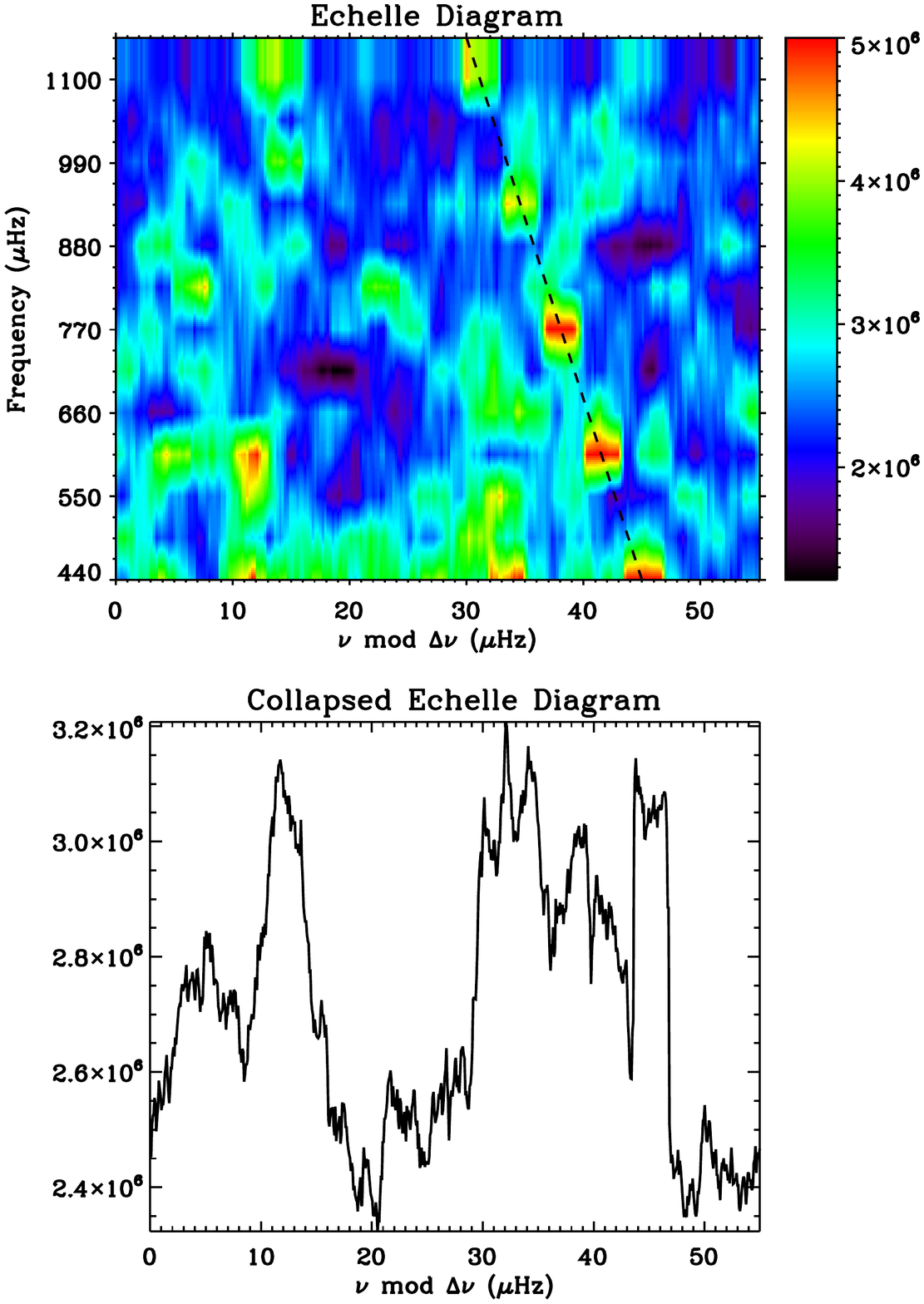}
\includegraphics[width=9.cm]{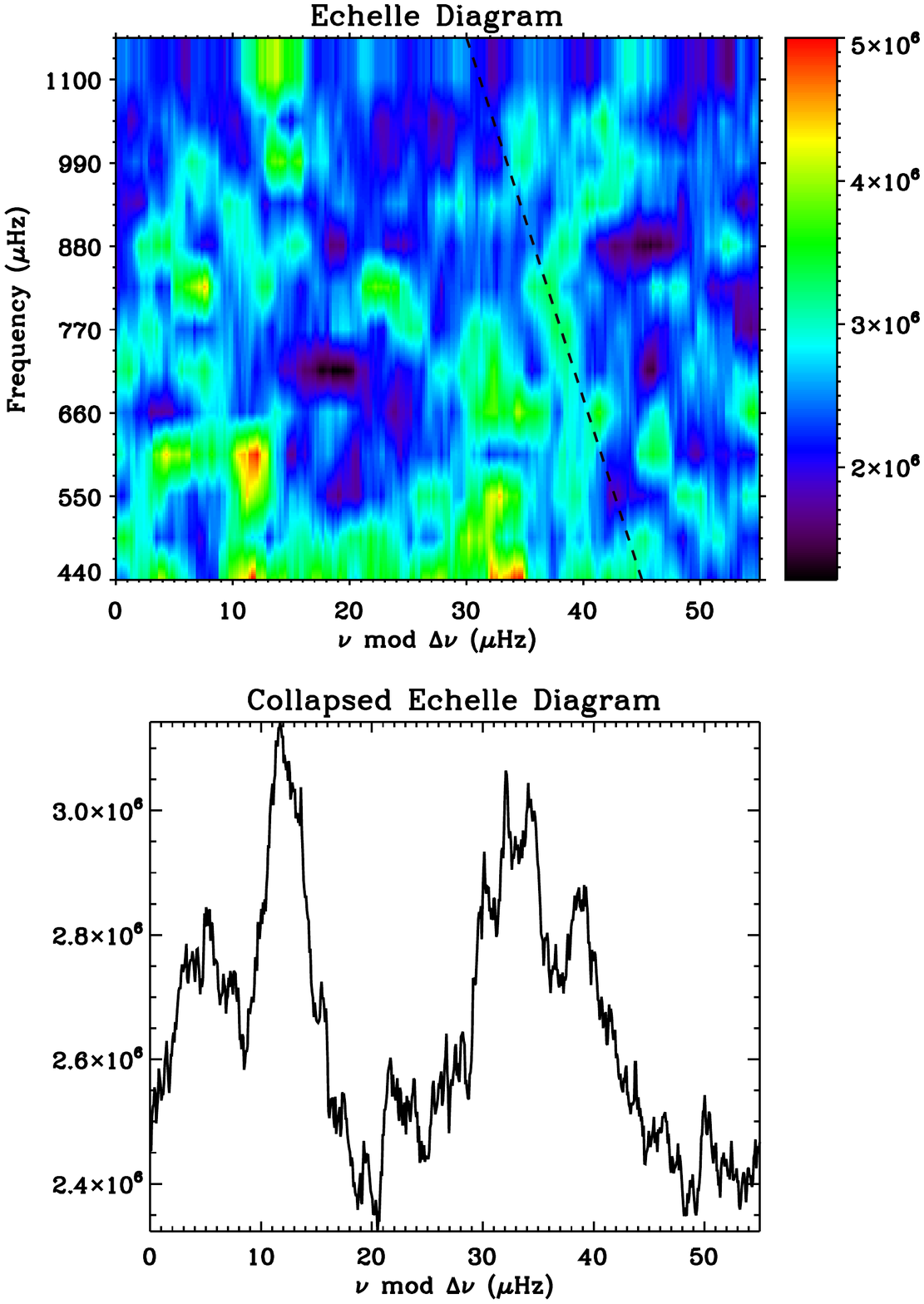}
\caption{Echelle diagrams obtained with the inpainted PSD (left panel) and the cleaned PSD (right panel) as explained in the text. They are computed using a folding frequency of 55 $\mu$Hz in the region [440 - 1200] $\mu$Hz where the excess p-mode power was found. Bottom panels are the integrated (collapsed) \'echelle diagrams along the vertical axis. The dashed line represents the places where the harmonics of the orbit are situated. }
\label{echelle_diag}
\end{figure*}

Let's note that a large separation around 55 $\mu$Hz is very close to the third harmonic of the orbit (53.9~$\mu$Hz) which are still present in the inpainted data. Therefore, we have repeated the analyses described before over a cleaned version of the power spectrum in which we have removed by hand the remaining peaks of the orbital harmonics substituting the affected bins by a local average of the PSD (see Fig.~\ref{echelle_diag}). The results remained unchanged which means that the pattern is not a consequence of an instrumental artifact. 

Using the large separation previously computed, we can calculate the so-called \'echelle diagram \citep{1983SoPh...82...55G}. To do so, we fold the power spectrum using a large separation of 55. $\mu$Hz in the region where the p-mode excess has been found [440 - 1200] $\mu$Hz. To reduce the dispersion of points in the spectrum we have first smoothed it by a boxcar function of 40 points (3.1 $\mu$Hz). The result is shown in the top, left panel of Fig.~\ref{echelle_diag}. This diagram is dominated by the presence of the harmonics of the orbit at multiples of 161.7 $\mu$Hz, which are seen as an inclined line of points starting at 48 $\mu$Hz. If we use the cleaned version of the PSD (where we have removed most of the peaks related to the orbital harmonics) and we do the same analysis (see top right panel in Fig.~\ref{echelle_diag}), two nearly vertical ridges clearly appear in the diagram. Indeed, if we integrate (collapse) the \'echelle diagram in the vertical direction we obtain two well separated ridges (see bottom panels in Fig.~\ref{echelle_diag}). We can tentatively assign the left-hand ridge with even-l modes and the right-hand one to the even ones odd knowing that the signal-to-noise ratio is not enough to go beyond that.

The values of the mean large separation found by the different groups agree within their error bars and in the following sections, we will take $\langle \Delta \nu \rangle\,=\,55.2\,\pm$\,0.8\,$\mu$Hz for the range [400 - 1200] $\mu$Hz. It is important to note that the error bar given here is the internal error of one single method and it does not take into account the dispersion in the results from the different pipelines and the slightly different ranges used in their computations.

\subsection{Variation of the large separation}

\begin{figure}
\centering
\includegraphics[width=9cm, trim=0.5cm 11cm 3cm 3.5cm]{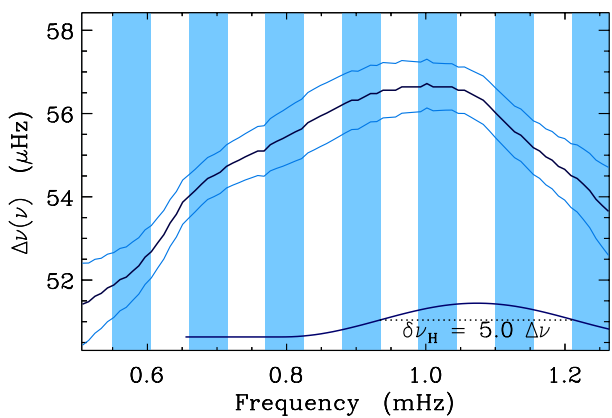}
\caption{Variation of the large separation with frequency. 1-$\sigma$ error bars are indicated. The extent of the Hanning filter used by the EACF is indicated in the lower-right corner, the dotted line measuring the full-width at half-maximum of the filter, here equal to 5 times the mean large separation. The width of the white and grey vertical regions is equal to $\dnumoy$.
\label{varidnuauto_mini}}
\end{figure}

When calculated with a filter narrower than the mode envelope, the EACF makes it possible to investigate the variation $\deltanunu$ of the large separation with frequency. With a filter width equal to 5 times the mean large separation, we test the variation at a large scale: the large separation increases with frequency at a rate $\ratednu$ of about 0.55\,$\pm$\,0.15\,$\mu$Hz per radial order, reaches its maximum at 980\,$\mu$Hz and then decreases (Fig.~\ref{varidnuauto_mini}).

\begin{figure}
\centering
\includegraphics[width=9cm, trim=0.5cm 11cm 3cm 3.5cm]{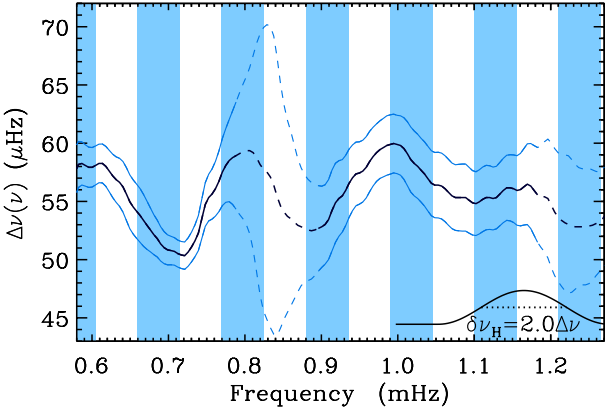}
\caption{Same as Fig.~\ref{varidnuauto_mini}, but with a narrow filter, corresponding to 2 times the mean large separation. The thin lines correspond to frequency ranges within the EACF is belove the threshold level for a rejection of the null hypothesis at the 1\,\% level. The dashed lines are the regions perturbed by the presence of orbital harmonics.
\label{varidnuauto_micro}}
\end{figure}

With a filter width equal to 2 times the mean large separation, we can investigate the rapid variation of the large separation. The number of independent measurements allows us to show the modulation of the large separation (Fig.~\ref{varidnuauto_micro}). For the study of the variation of the large separation with frequency $\deltanunu$, the threshold level is lower than the one for the mean large separation, benefitting from the fact that the large separation is then searched in a pre-determined reduced interval compared to the blind analysis. Most of the values of the EACF with such a filter are above 4.6, which corresponds to the 1\,\% rejection level for the analysis with a 2-$\dnumoy$ wide filter  and a search in the frequency range allowing 20\,\% variation with respect to the mean large separation. They give error bars lower than 1\,$\mu$Hz, except in the ranges [790 - 900]\,$\mu$Hz  and [1180 - 1280]\,$\mu$Hz. The EACF values are large enough to enable an unambiguous detection of the modulation. The period $W$ of this modulation, close to 200\,$\mu$Hz, could be caused primarily by the He\,II ionisation zone or it could be related to the internal phase shifts.
Following \cite{2005MNRAS.361.1187M}, we infer the acoustic depth $\tau \simeq 1/2W$ of this region in the range $\tau=2\,300$--$2\,700$\,s.

\begin{figure}[h]
\centering
\includegraphics[width=9cm, trim=0.5cm 11cm 3cm 3.5cm]{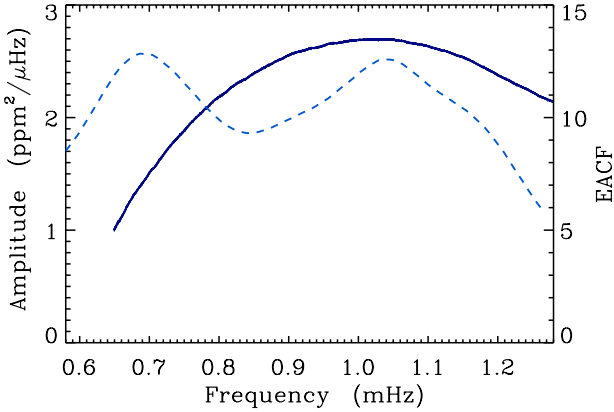}
\caption{Comparison of the EACF (dashed line), estimated with a 5-$\dnumoy$ large filter, and the bolometric mode amplitude expressed in ppm (continuous line).
\label{comp_amp_ampl}}
\end{figure}

The analysis of the variation of the large separation with frequency, $\deltanunu$, shows a peculiar behavior in the range [790 - 900]\,$\mu$Hz. One would expect large values of the EACF in this range close to $\numax$, which is not the case. Since the variations of $\deltanunu$ are not greater in this region than in others, one can explain the low values of the EACF either by a perturbation caused by a mixed mode or by a low value of the mode lifetime.
The presence of a mixed mode is in fact quite unlikely: such a mode with a much longer lifetime than normal p mode should appear with a large amplitude, which is not the case. Thus, one has to favor the hypothesis of the varying mode lifetime. The comparison of the EACF with the bolometric amplitude per radial mode (see Sect.~\ref{secampl}) indicates a strong increase of the mode lifetime below 850\,$\mu$Hz and a strong decrease above 1100\,$\mu$Hz (Fig.~\ref{comp_amp_ampl}). This behaviour is in agreement with the measurement made in a similar star with a very close large separation, such as HD 49385 \citep{2010arXiv1003.4368D}, in spite of a different effective temperature. However, it is not yet possible to quantify the lifetime.


\subsection{Maximum bolometric amplitude}\label{secampl}

To compute the maximum bolometric amplitude per radial mode we need to smooth the p-mode hump and correct from the background in this frequency range. To do so, we have used different methods \citep[see for an extensive explanation][]{2008ApJ...682.1370K,2010MNRAS.402.2049H,2010A&A...511A..46M}. Then to correct from the instrumental response function we use the method developped by \citet{2009A&A...495..979M}. Thus, we have obtained a bolometric amplitude per radial mode of A$_{\rm{bol}, l=0}$\,=\,2.7\,$\pm$\,0.2\,ppm at $\sim$1070~$\mu$Hz. However, among the different teams and depending on the method and on the way we fit the background, the value for the maximum amplitude per radial mode varied within the range 2.4 to 2.9~ppm indicating that the error bar of 0.2 ppm could be underestimated. So we have estimated a new value for the error of 0.6~ppm, which is derived from the scatter of the smoothed power spectrum about the background fit outside the oscillation range \citep{2009CoAst.160...74H}.

\begin{figure}[!htbp]
\begin{center}
\includegraphics[width=9cm]{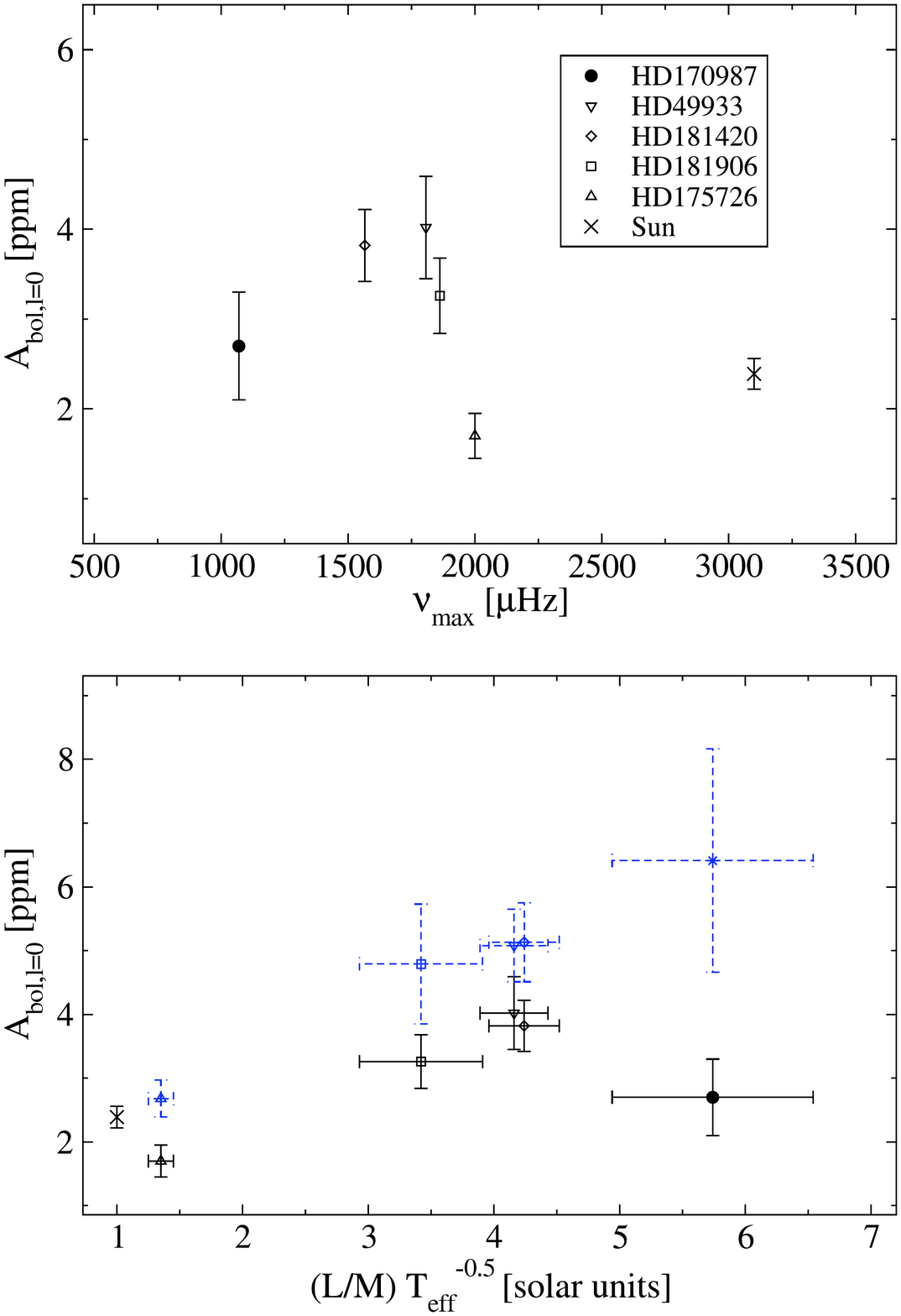}
\caption{Top panel: Maximum bolometric amplitude per radial mode versus $\nu_{\rm max}$ for published CoRoT F-stars 
(open symbols), the Sun (star symbol) and HD 170987 (filled circle).  Bottom panel: Maximum bolometric amplitude per radial mode versus stellar parameters. Black 
symbols are the observed values, and blue symbols theoretical values using  $(L/M)^{0.7}$. The symbol types are the 
same as in the top panel.}
\label{abol}
\end{center}
\end{figure}

The top panel of Fig.~\ref{abol}  shows the comparison of the amplitude found for HD~170987 with the published values of 
four other CoRoT solar-like targets \citep{2008Sci...322..558M, 2009A&A...506...33M} as a function of $\nu_{\rm max}$. The bottom panel shows the amplitudes compared to theoretical values for each star calculated using the $(L/M)^{s} T_{\rm eff}^{-0.5}$ scaling relation from Eq.~3 of \citet{kjeldsen95} with s\,=\,0.7 \citep{2007A&A...463..297S}. It can be seen that while in all cases the amplitudes are systematically about 30\,\% lower than the theoretical values, the discrepancy appears to be significantly larger for HD~170987 ($\sim$ 50\,\%). We note that the error bars are larger for HD~170987 due to the rather large uncertainty on the parallax ($\sim$10\,\%). In both cases, the discrepancy is well inside 3-$\sigma$.


 

\section{Inferring global stellar parameters}

Using $\langle \Delta \nu \rangle = 55.2\,\pm\,0.8\,\mu$Hz, 
$f_{\rm min} = 400\,\mu$Hz, $f_{\rm max} = 1200\,\mu$Hz (see Sect.~5.1 and 5.2), and 
the spectroscopic information 
$\log g = 4.20\,\pm\,0.14$~dex, 
$[M/H] = -0.20\,\pm\,0.15$~dex and $T_{\rm eff} = 6540\,\pm\,80$~K 
(see Sect.~\ref{secspec}), we
compare these values with stellar models to determine the 
radius and the mass of HD~170987.
The stellar models that we use are the 
Aarhus Stellar Evolution Code (ASTEC) 
coupled with an 
adiabatic pulsation code (ADIPLS) (Christensen-Dalsgaard 2008a,b).  
These codes need as input the stellar parameters of mass, age, chemical
composition, and mixing-length parameter, 
and return stellar observables $B_i$, such as radius,  
effective temperature, and 
the frequencies of the oscillation modes.

The parameters that best describe the observables are obtained by 
minimizing a $\chi^2$ function; 
\begin{equation}
\chi^2 = \sum^M_{i=1} \left ( \frac{y_i - B_i}{\epsilon_i}\right )^2,
\end{equation}
where $y_i$ and $\epsilon_i$ are the $i=1,2,...,M$ observations and errors. Here, we have $M$\,=\,4.
The Levenberg-Marquardt algorithm is used for the optimization, and this
incorporates derivative information to guess the next set of parameters
that will reduce the value of $\chi^2$.  Naturally, an initial guess
of the parameters is needed and these are obtained from a small grid 
of stellar evolution tracks.

Because there are few observations and just as many parameters, there are
inherent correlations between mass, age and chemical composition. 
To help avoid local minima problems, we minimize the $\chi^2$ function 
beginning at several initial guesses of the parameters (mainly varying
in mass and age), and these initial guesses are estimated from the grids. 
We therefore obtain several sets of parameters with a 
corresponding $\chi^2$ value that match the observations as best as possible.
Choosing the models whose $\chi^2$ values fall below 3.9$^2$ (assuming that the errors are Gaussian and that we want a 99\,\% probability), 
the radius and mass are determined
by calculating the average value of each, and the uncertainty is given by
the dispersion in the values divided by 6 
\citep[see][for details]{2010A&A...511A..46M}.
We estimate the age of the star by simply matching the ``fitted mass'' 
with the closest age from the model parameter results, we therefore have no
quantitative measure of its uncertainty. leading to a very preliminary value. 
The results are given in Table~\ref{tab:rm}.  The estimate of the age
implies a star coming close to the end of the main sequence phase.

\begin{table}
\begin{center}
\caption{Stellar parameters from modelling\label{tab:rm}}
\begin{tabular}{lll}
\hline\hline
Parameter & Value & Error\\
\hline
Radius ($R_{\odot}$) & 1.960 & 0.046\\
Mass ($M_{\odot}$)& 1.43 & 0.05 \\
Age (Gyr) & 2.4 & --\\ 
\hline\hline
 \end{tabular}
\end{center}
\end{table}

\section{Discussion}


Thanks to spectroscopic observations from NARVAL, we have been able to estimate the effective temperature as well as the chemical abundances of HD~170987 showing that this star has a low metallicity (-0.15\,$\pm$\,0.06\,dex) and a high abundance of lithium.

The surface light elements abundances could give us
an interesting clue about the internal rotation of HD~170987.
The effective temperature of $\rm 6540\,K$ sets this star in the so-called
lithium dip where both $\rm ^7Li$ and $\rm ^9Be$ are depleted \citep{2002ApJ...565..587B}.
This abundance dip is most likely related to
rotation and observations show that depletion
and rotation are correlated \citep{1987PASP...99.1067B}.
\citet{2003A&A...405.1025T} explained the cooler side
of the lithium dip (from 6500 to 6200 K) by invoking an increasing efficient transport
of the internal angular momentum through gravity waves when
the mass decreases from $\approx 1.4$ to $\approx 1.2\,M_{\odot}$.
For the more massive stars differential rotation in
the interior leads to shear instabilities: the radiative
envelope is mixed deeply and light elements are destroyed
through proton capture. For the less massive stars there is less
differential rotation and less shear instabilities. Thus
no (or far less) depletion occurs.

The lithium dip concerns stars that have an effective temperature in the range 6400 to 6850\,K. At Hyades age this star should already have exhibited
a depletion of about 0.5 dex in $\rm^7Li$ \citep{2002ApJ...565..587B}.
However HD~170987 seems much older than Hyades
and shows no such depletion: if we combine the current solar $\rm ^7Li$
abundance \citep{2009arXiv0909.0948A} and the relative abundance
of HD170987 to the Sun given above we obtain $\rm A(^7Li) \approx 2.9$~dex.
This value is very close to the fraction of lithium
in the material out of which Population I stars form: 3.3 to 3.2 dex. Thus the abundance of $\rm^7Li$ in the atmosphere of HD~170987
directly supports the absence of deep mixing and
indirectly supports the absence of differential rotation with
depth. At least if some internal differential rotation exists it is not sufficient to induce
deep mixing.


As confirmed by our analysis, HD~170987 is remarkably similar to Procyon, a prime target for previous asteroseismic studies using ground-based spectroscopy and space-based photometry.  For the latter, observations by the MOST satellite resulted in a null-detection of p-modes \citep{MatKus04}, leading to conclusions that were later critically discussed by \citet{2005A&A...432L..43B}. These observations 
were followed by WIRE photometry \citep{BruKje2005} and higher precision MOST data collected in 2007 \citep{2008ApJ...687.1448G}, both showing power excess around the expected p-mode location of 1mHz, while velocity observations revealed a clear p-mode structure \citep[see, e.g.,][]{2008ApJ...687.1180A,2010ApJ...713..935B}. Convection models by \citet{2008ApJ...687.1448G} showed that for Procyon, as opposed to the Sun, the granulation timescale can coincide with the location of pulsation frequencies. The resulting interaction with the acoustic 
oscillations causing short mode lifetimes were identified as possible difficulties for detecting clear peak spacings in the photometry of Procyon-like stars. Given the similarity of HD~170987 to Procyon, it is tempting to speculate that a similar mechanism is in part responsible for the low visibility of the pulsation modes in the CoRoT photometry of HD~170987. Alternate mechanisms which could partly explain the low signal include the low metallicity \citep[see][]{2009arXiv0910.4027S} or the fact that some of the light is diluted by a possible binary companion.

\section{Conclusions}

We have analysed the 149~days of the light curve of the CoRoT target HD~170987, allowing us to determine its global parameters. We find that the rotation period of the star is around 4.3 days and the v $\sin i$ of 19\,km/s, leading to an inclination angle $i$ of $\sim$\,50$\degr \;^{+20}_{-13}$. By fitting the background, we obtain that the time-scale for the granulation is around 383~$\pm$\,28~s either assuming that we have some contribution of faculae or of modes.

We confirm that power excess between 400 and 1200~$\mu$Hz is due to acoustic modes, which are characterised by a mean large separation $\langle \Delta\nu \rangle$~=~55.2~$\pm$\,0.8~$\mu$Hz with more than 95\,\% of probability and a posteriori probability of 68\,\%. Using the EACF method, we find  $\langle \Delta\nu \rangle$~=~55.9~$\pm$\,0.2\,$\mu$Hz in a higher frequency range [500 - 1250] $\mu$Hz with a 1\,\% rejection. However, we recognize that all these confidence levels are dependent on our assumption that our spectrum is dominated by white noise statistics. It may therefore be that we should be less confident as there could be other sources of noise embedded in the signal. However, at such a level, the detection seems to be unambiguous. Moreover, this value is in agreement with the expected one derived from the stellar fundamental parameters, which gives 52.7\,$\pm$\,4\,$\mu$Hz, the large error bar being due to the large uncertainty on the parallax. Now, thanks to the detection of this pattern, we can be confident that the bump observed around 1000~$\mu$Hz is very likely to be related to acoustic modes.

We characterise the power excess with a maximum amplitude of 2.7~$\pm$\,0.6~ppm at $\sim$1070~$\mu$Hz, where the uncertainty on the amplitude has been obtained by taking into account the scatter of the smoothed power spectrum about the background fit  outside the oscillation range.

Because of the low signal-to-noise ratio, we cannot fit the acoustic modes individually. However, from the global seismic parameters found and the results form the NARVAL spectrohraph, we have estimated that HD~170987 has a mass, M~=~1.43~$\pm$\,0.05~$M_\odot$, a radius, R~=~1.96~$\pm$\,0.046~$R_\odot$, and an age $\sim$2.4~Gyr. 

Further studies are needed to have a better comprehension of the propagation of waves in stellar atmosphere of stars located at this position in the HR diagram. It would also enable us to understand why the oscillation-amplitudes measured in intensity fluctuations are so small.


\begin{acknowledgements}
The authors want to thank S. Pires and J.L. Stark for their useful comments and discussions on the inpaining methods applied in this paper as well as T.R. Bedding for his useful comments. D.S. acknowledges Australian Research Council. J.B. acknowledges the ANR Siroco of the French agency for research. This work has been partially supported by: the CNES/GOLF grant at the Service d'Astrophysique (CEA/Saclay) and the grant PNAyA2007-62650 from the Spanish National Research Plan. This work benefited from the support of the International Space Science Institute (ISSI), through the workshop programme award. It was also partly supported by the European Helio- and Asteroseismology Network (HELAS), a major international collaboration funded by the European Commission's
Sixth Framework Programme. IWR, GAV, SH, WJC, and YE acknowledge support from the UK  Science and Technology Facilities Council (STFC).  

\end{acknowledgements}

\bibliographystyle{aa} 
\bibliography{/Users/Savita/Documents/BIBLIO_sav}  

\appendix
\section{Inpainting interpolation of the light curve}

``Inpainting" refers to methods that interpolate the missing information using some priors on the solution. The method that we applied to HD 170987 uses a prior of sparsity and was introduced by Elad et al. (2005). It has already been applied to various fields of astrophysics, such as weak lensing \citep{2009MNRAS.395.1265P} and recently to asteroseismic time series by Sato et al. (in prep.). It assumes that there exists a dictionary $\Phi$ -- which in our case is a Discrete Cosine Transform (DCT) -- where the complete data are sparse and the incomplete data are less sparse. It means that there exists a representation $\alpha = \Phi^TX$ of the signal $X$ in the dictionary $\Phi$ where most coefficients $\alpha_i$ are close to zero.
Let $X$ be the ideal complete time series, $Y$ the observed time series and $M$ the window function or mask (i.e. $M_i=1$ where there is a valid data point and $M_i=0$ elsewhere). Thus we have that $Y=MX$. Inpainting consists in recovering $X$ knowing $Y$ and $M$. 
The solution is obtained by minimizing the following equation:
\begin{equation}
\min_{X}  \| \Phi^T X \|_1    \textrm{ subject to } \sum_i ( Y - MX )^2 \le \sigma,
\end{equation}
where $\sigma$ stands for the noise standard deviation and where we use a pseudo norm, with $ \| z \|_1 = \sum_k \mid z_k \mid $. To minimize this expression, an iterative algorithm was used. 

In this study, we have built the mask $M$ to remove those points which are affected by the SAA and also all the other points that were already linearly interpolated, according to the {\it status} flag provided in the N2 files. 

Most of the gaps in the CoRoT time series have a typical time scale of several minutes. It is possible to interpolate these typical gaps by using an inpainting algorithm based on a DCT decomposition which is localized in a short time scale. However, there are also a few longer gaps which have a time scale of several hours in the time series. The inpainting algorithm needs to take into account this large variation of the scale of the gaps. To do that, we have replaced the simple DCT decomposition by a Multi-Scale Discrete Cosine Transform decomposition. With this transform, we first decompose the light curve into different scales using a wavelet transform. Then, the inpainting of each scale makes possible to interpolate the gaps of different time scales. The PSD of the inpainted time series is shown in Fig~\ref{psdfull} (bottom). The improvement is clearly visible by a reduction of the amplitude of the orbital harmonics by more than 90\,\% in many cases, as well as the disappearance  of the daily modulation except for the first orbital harmonic (see the inset of Fig~\ref{psdfull} (bottom) corresponding to the second orbital harmonic in which the main peak is reduced by 95\,\% and no signature of the daily modulation is seen).

\end{document}